\documentstyle[12pt]{article}
\catcode`\@=11
\begin{document} \openup6pt

\title{VISCOUS COSMOLOGIES WITH EXTRA DIMENSIONS}

\author{Bikash Chandra Paul\thanks{Electronic mail : bcpaul@nbu.ernet.in}\\
	Physics Department, North Bengal University, \\
	Siliguri, Dist. Darjeeling, PIN 734 430, India}

\date{}

\maketitle

\vspace{0.5in}

\begin{abstract}

We present an analysis of a n-dimensional vacuum Einstein field equations
in which 4-dimensional space-time  is described by a Friedmann 
Robertson-Walker (FRW) metric and that of the extra dimensions by a Kasner
type Euclidean metric. The field equations are interpreted as four 
dimensional 
Einstein equations with effective matter properties.
The effective matter is then treated as viscous
fluid. We consider the theories of imperfect fluid given by Eckart,
 truncated Israel-Stewart  and full Israel-Stewart theories and obtain
cosmological solutions  for a flat model of the universe.
The imperfect fluid assumption admits here power law inflation for the
 3-physical space in some cases. 

\end{abstract}


\pagebreak

Model building in higher dimensions was initiated by Kaluza and Klein$^{1}$
who tried to unify gravity with electromagnetic interaction by introducing an 
extra dimension. Kaluza-Klein theory is basically an extension of Einstein general 
relativity in 5D which is of much interest in Particle Physics and Cosmology.
In the last few decades the study of a higher dimensional theory has been 
revived and considerably generalized after realizing that many interesting
theories of particle interactions need more than four dimensions for their 
formulation. Attempts have been made to build cosmological models$^{2}$ 
in higher dimensions which may undergo a spontaneous compactification 
leading to a product space $M^{4} \times M^{d}$, with $M^{d}$ describing the 
compact  ${\it inner \;  space }$. In the usual approach one uses Einstein's field equation
\begin{equation}
G_{ab} = 8 \pi G \;  T_{ab} 
\end{equation}
which match the Einstein tensor and the energy momentum tensor.
Recently another approach$^{3}$ has been developed taking $G_{ab} = 0$, the 
extra terms which appear in these equations due to extra dimensions are 
interpreted as induced or effective properties of matter in ordinary 4D
space-time. Ib\'{a}nez and Verdaguer [henceforth IV]$^{4}$
 obtained a set of solutions  of
Einstein's equation in an n-dimensional vacuum. They found homogeneous solutions
with expanding three dimensional isotropic space. 
The solutions are then
identified with the observable four dimensional subspace with perfect fluid.
Later a class of  solutions in flat universe model have been
 obtained by Gleiser and Diaz.$^{5}$ Krori 
{\it et al.}$^{6}$ using a higher dimensional anisotropic cosmology
 (Bianchi-I)  obtained 4D
 perfect fluid solutions. It was shown that the
 perfect fluid solutions obtained by them are 
compatible with contraction of all the extra dimensions.
The solutions are identified with a 4 dimensional anisotropic cosmological
model. However, it is known that the matter distribution in the 
early universe may not be as simple as predicted by perfect fluid due
to a number of dissipative processes.  Hence it is essential to study
cosmological solutions in the framework of imperfect fluid description. The
simple theory is the Eckart theory,$^{7}$ however, it suffers from serious drawbacks concerning
causality and stability.$^{8,9}$   These difficulties can, however, be removed
by considering higher order theories i.e., extended irreversible thermodynamics 
(EIT).$^{10,11}$
In this paper we intend to study a 
higher dimensional vacuum Einstein equations which leads to an  observed
4 dimensional universe with imperfect fluid. 
 We obtain exact solutions
in the framework of the Eckart, truncated Israel-Stewart (TIS) and full Israel-Stewart
 (FIS) theories.

We now consider a higher dimensional metric in the form
\begin{equation}
ds^{2} =  ds^{2}_{FRW}  + \sum_{i = 1}^{d} b_{i}^{2} (t) \; dx_{i}^{2}
\end{equation}
where $d$ is the number of extra dimensions $( d = n - 4 )$
and  $  ds^{2}_{FRW} $ represents  the line element of the FRW metric
in four dimensions which is given by
\begin{equation}
ds_{FRW}^{2} = - \; dt^{2} + a^{2} (t) \left[ \frac{dr^{2}}{1 - kr^{2}} + 
r^{2} \left( d \theta^{2} + sin^{2} \theta \; d \phi^{2} \right) \right]
\end{equation}
where $ a(t)$ is the scale factor  of the 4 dimensional spacetime and $b_{i}$'s are 
the scale factors for the extra dimensions and $ k = 0, + 1, -1 $ represents
flat, closed and open universe respectively. 

The vacuum Einstein's equation in n-dimensions using the metric (2)
 can be written as
\begin{equation} 
3 \left( \frac{\dot{a}^{2} + k}{a^{2}} \right) = \frac{\ddot{D}}{D} 
   - \frac{1}{8} \left( \sum_{i = 1}^{d} 
       \frac{\dot{b}_{i}}{b_{i}} \right)^{2}
+ \frac{1}{8} \sum_{i = 1}^{d} 
       \left( \frac{\dot{b}_{i}}{b_{i}} \right)^{2} ,
\end{equation}
\begin{equation} 
2 \frac{\ddot{a}}{a} + \frac{\dot{a}^{2} + k}{a^{2}} = 
        \frac{\dot{a} \dot{D}}{a D} 
     +  \frac{1}{8} \left( \sum_{i = 1}^{d} 
       \frac{\dot{b}_{i} }{b_{i}} \right)^{2}
- \frac{1}{8} \sum_{i = 1}^{d} 
       \left( \frac{\dot{b}_{i}}{b_{i}} \right)^{2} ,
\end{equation}
\begin{equation} 
\frac{ \ddot{b_{i}}}{b_{i}} + 3 \frac{\dot{a} \dot{b}_{i}}{a b_{i}} 
      +  \frac{\dot{D} \dot{b}_{i}}{D b_{i}} -  
\frac{\dot{b_{i}}^{2} }{b_{i}^{2}} 
    = 0 
\end{equation}
where  we denote $ D^{2} = \Pi_{i = 1}^{d}  b_{i} (t)$.

The  field 
eqs. (4)-(6) can be rewritten as the Einstein equations with imperfect fluid
in 4 dimensions. Using  $b_{i} = D^{2 p_{i}}$ with
 $\sum_{i = 1}^{d} p_{i} = 1 $  the field equations reduce to
\begin{equation} 
3 \left( \frac{\dot{a}^{2} + k}{a^{2}} \right) = \rho
\end{equation}
\begin{equation} 
2 \frac{\ddot{a}}{a} + \frac{\dot{a}^{2} + k}{a^{2}} = - P_{eff}
\end{equation} 
where 
$\rho = \frac{\ddot{D}}{D} 
   + \alpha \left( \frac{\dot{D}}{D} \right)^{2}$ ,  
$P_{eff} = p +  \Pi = - \frac{\dot{a} \dot{D}}{a D} + \alpha
 \left( \frac{\dot{D}}{D} \right)^{2}$ 
and    
 $\alpha = \frac{1}{2} \left[ \sum_{i = 1}^{d} p_{i}^{2} - 1 \right]$. 
In the above
 $p$ represents the thermodynamic pressure and
 $ \Pi$ represents  the bulk viscous stress. Using the relation $p = (\gamma - 1) 
\rho   \; \; ( 1 \leq \gamma \leq 2 )$ one obtains the bulk viscous
stress which
is given by
\begin{equation} 
\Pi = - \left( \gamma - \frac{4}{3} \right) \frac{\ddot{D}}{D} 
         -  \alpha (\gamma - 2) \left( \frac{\dot{D}}{D} \right)^{2} .
\end{equation}
It is evident above that for $\alpha = 0$ and $\gamma = \frac{4}{3}$ one recovers
the cosmological solutions for a radiation dominated universe 
i.e. without viscosity. The solutions are obtained by IV in the context of perfect
fluid in 4 dimensions.
We study
 here cosmological
solutions for $\alpha \neq 0$, which leads to interesting results. 
 The evolution of the bulk
viscous stress in (3 + 1) dimensions is given by$^{10}$
\begin{equation} 
\Pi + \tau \dot{\Pi} = - 3 \zeta H - \frac{\epsilon}{2} \tau \Pi
\left( 3 H +  \frac{ \dot{\tau}}{\tau} - \frac{ \dot{\zeta}}{\zeta} -
\frac{\dot{T}}{T} \right) ,
\end{equation}
where $\zeta $ is the coefficient of bulk viscosity, $\tau$ is the relaxation time
and $ T $ is the temperature and H ( = $ \frac{\dot{a}}{a} $ ) 
 
is the Hubble parameter. Here the parameter $\epsilon $ can take the value
either 0 or 1;  $\epsilon = 0$ for TIS and $\epsilon = 1 $ for FIS theory. One obtains
Eckart theory for $\tau = 0 $. The system of eqs. (7)-(10)
are not closed  as number of unknowns are more than the number of equations. 
It is, therefore, necessary to assume the following adhoc but commonly
chosen relations for the bulk viscosity co-efficient and relaxation time :
\begin{equation}
\zeta = \beta \rho^{q} \; \;   and \; \; \tau = \beta \rho^{q-1}
\end {equation}
where $q \geq 0$ and $\beta $ is a constant parameter.  The behavior of  temperature
 $ T $  of a universe with imperfect fluid is obtained in FIS
theory using eq.(10).

We now look for cosmological solutions. Let us consider a power 
law model of the expansion of the universe given by
\begin{equation}
a = a_{o} t^{m} 
\end {equation}
where $ a_{o} $ and m are constants. 
We consider a flat universe ( $k = 0$ ) and 
obtain solutions for Eckart, TIS and FIS theories for two cases 
$\gamma = \frac{4}{3}$ and $\gamma = 2$ respectively.

Case I : Eckart theory :

Eckart theory corresponds to $\tau = 0$. The eq.(10)
determines the viscous stress, using
$q = \frac{1}{2}$ i.e., $\zeta = \beta \rho^{1/2}$ we get the 
following solutions :
 
 (i)  For $\gamma = \frac{4}{3}$
one gets power law expansion with  $\beta = \frac{2 (2 m - 1)}{\sqrt{3} m} $.
The coefficient of bulk viscosity $\beta $ is a positive quantity 
  for 
$ m > \frac{1}{2}$. The model admits a large power law inflation
($m \rightarrow \infty $)   when  $\beta \rightarrow 
\frac{4 }{\sqrt{3}} $.

(ii) For $\gamma = 2$, we get 
$\beta = \frac{2 (3 m - 1)}{3 \sqrt{3}m}$. In this regime a  large power 
law inflation is obtained when $\beta \rightarrow \frac{2}{\sqrt{3}}$.

Case II :  TIS theory ( $\epsilon = 0$ ) :

(i)   For $ \gamma = \frac{4}{3} $, eq.(9) determines the bulk viscous stress
which is given by 
\begin{equation} 
\Pi =  \frac{2 \alpha (1 - 3 m)^{2}}{3 t^{2}}. 
\end{equation}
The bulk viscous stress depends both on $\alpha$ and on the exponent $m$. 
For $m = \frac{1}{3}$ , the co-efficient of bulk viscosity vanishes.
 Thus one requires $\alpha < 0$ and  $m \neq \frac{1}{3}$ 
for an acceptable solution. The first constraint in the model is
satisfied for $\sum_{i=1}^{d} p_{i}^{2} < 1 $.
 For simplicity we consider $q = \frac{1}{2}$ and obtain
$ \beta = \frac{ 2 \sqrt{3} (2m - 1) m}{9m^{2} + 8 m - 4} $ from eq.(10).
In this case a realistic scenario is obtained for  $m > \frac{1}{2}$. One
gets a large power law inflation when $\beta \rightarrow \frac{4 \sqrt{3}}{9}$.

(ii)   For $ \gamma = 2 $,  the bulk viscous stress is
given by 
\begin{equation} 
\Pi =  - \; \frac{2 m (3 m - 1)}{ t^{2}} ,
\end{equation}
which remains negative for $m > \frac{1}{3}$. The 
bulk viscous stress is 
independent of $\alpha$.
 For $q = \frac{1}{2}$, we get
$\beta = \frac{ 2 \sqrt{3} m (3m - 1)}{ 9 m^{2} + 12 m - 4 }$, which allows
 a realistic scenario for 
$ m > \frac{1}{3}$. In this case a large power law expansion is derived
when $\beta \rightarrow \frac{2 \sqrt{3}}{3}$.

Case III : FIS theory ( $ \epsilon = 1$ ) :

 In this case temperature of the universe can be evaluated from the eq.(10).

(i)
For  radiation ( $\gamma = \frac{4}{3}$ ) and viscous fluid model,
the bulk viscous stress is evaluated which is given by
\begin{equation} 
\Pi = \frac{2 \alpha}{3} \frac{(1 - 3m )^{2}}{t^{2}} .
\end{equation}
For simplicity we take  $q = \frac{1}{2}$ and 
obtain the temperature of the universe
which is given by 
\begin{equation} 
T = T_{01} t^{a_{1}}
\end{equation}
where $ a_{1} = 3 m - 4 + \frac{2 \sqrt{3} m}{\beta} + 
\frac{27 m^{3}}{\alpha
(1 - 3m)^{2}} $ and $T_{01} $ is an integration constant.
One gets a decreasing mode of temperature in this regime for
$\beta > \frac{2 \sqrt{3} m ( 2m - 1)}{3 m^2 + 11 m - 4}$,

 which in turn determines the relaxation time and coefficient of bulk 
viscosity. Thus a lower bound on viscosity is obtained depending on the 
exponent of the scale factor.
A realistic scenario is obtained  for $m  \geq
 \frac{1}{2}$. However, for a large power law inflation one requires a large value
of the bulk viscosity (as  $\beta > \frac{4 \sqrt{3}}{3}$).

(ii) For a stiff matter ( $\gamma = 2$ ) and viscous fluid model, 
the bulk viscous stress is 
\begin{equation} 
\Pi = - \frac{ 2 m (3m - 1)}{t^{2}}  ,
\end{equation}
which is independent of $\alpha$. Here $\Pi < 0$ is obtained for
 $ m > \frac{1}{3}$. Using $q = \frac{1}{2}$  one gets temperature
of the universe which varies as  
\begin{equation} 
T = T_{02} t^{a_{2}}
\end{equation}
where $ a_{2} =  \frac{4 - 15 m}{3 m - 1} + \frac{2 \sqrt{3}
 m}{\beta}$ and $T_{02}$ is an integration constant.
 In this case a decreasing mode of temperature is found for
$\beta > \frac{2 \sqrt{3} m (3m - 1)}{15 m - 4} $. For a realistic scenario
 one requires    $m \geq \frac{1}{3}$ in this case. One gets a large power law expansion
for $\beta > \frac{2 \sqrt{3}}{5}$.

Thus we obtain cosmological solutions of a higher dimensional universe
 considering matter sector of the theory which arises from the extra 
dimensions. The evolution of the extra  dimensions scale factor
 is described by Kasner-type behavior. The solutions
obtained here are identified with the (3 + 1) dimensional flat (k = 0)
universe with imperfect
fluid. The solutions are obtained in Eckart, TIS and FIS theories respectively.
The causal fluid distribution considered here admit cosmological models 
with power law inflation as well as  models with slow power law expansion 
of the early universe.  The overall evolution of the internal space is contracting in nature 
( as $ \sum_{i = 1}^{d} p_{i}^{2} < 1 $ and $ \sum_{i = 1 }^{d}  p_{i} = 1 $ ) 
whereas the 3-physical space expands. We note the following :

$\bullet $
 In the TIS or FIS theory one may get a universe with comparatively large  power law expansion 
when $\gamma = \frac{4}{3}$ than that when $\gamma = 2$ for a given bulk 
viscosity.  However  
for $\gamma = \frac{4}{3}$ with $m = \frac{1}{2}$ one gets back the solution 
corresponding to perfect fluid distribution. 
The usual radiation dominated  universe
 $ \left( a(t) \sim \sqrt{t} \right) $ is recovered in the case.$^{4}$ 

$\bullet $
A large viscosity drives a power law inflation. In this model
the viscosity parameter $\beta$ is least in TIS theory and greatest in FIS theory
for a given expansion of the universe for $\gamma = \frac{4}{3}$ as well as
for $\gamma = 2$. Thus in the TIS theory, one gets a large power law inflation for
a comparatively lower bulk viscosity.
However one gets a range of values 
of $\beta$ in FIS theories which requires a higher bulk viscosity
 to begin with  for $\gamma = \frac{4}{3}$ compared to the other models.
  The viscosity of the
universe decreases as the universe evolves.

$\bullet$
The temperature of the universe can be determined explicitly in the FIS theory which
is found to have a decreasing mode
 determined by the viscous fluid distribution.
The bulk viscosity of the universe decreases with time with the choice 
$q = \frac{1}{2}$ considered here in eq. (11).
 One interesting aspect of the solution is that for a given expansion 
of the universe $\beta$ is fixed in Eckart and TIS theories with values
 $\beta (Eckart) > \beta (TIS)$. However one finds a lower bound on the value
of viscosity in the FIS theory. In this case 
a hot universe cools rapidly when $\beta $ i.e., viscosity  
of the universe is large. 
 
\vspace{0.2in}

{\bf  \it Acknowledgements :}

I would like to  thank the Inter-University Centre
for Astronomy and Astrophysics ( IUCAA ) Pune for awarding a Visiting Associateship and providing a facility where this work was carried out.  I am grateful to Professor N. Dadhich for his suggestion and cordial discussion.
 Finally, I like to thank Professor S. Mukherjee for providing 
${\it IUCAA \; Reference \;  Centre}$ (IRC) facilities at North 
Bengal University for completing the work.

\pagebreak

\end{document}